\documentclass[superscriptaddress,aps,prl,twocolumn,showpacs,preprintnumbers,amsmath,amssymb]{revtex4}

\usepackage{graphicx}% Include figure files
\usepackage{dcolumn}% Align table columns on decimal point
\usepackage{bm}% bold math

\begin{document}

\preprint{Submitted to Phys.~Rev.~Lett.(2010)}
\title{Broadband Slow Light Metamaterial Based on a Double-Continuum Fano Resonance}
%\title{Slow Light without Electromagnetically Induced
%Transparency: the Case of a Double Fano Resonance}

\author{Chihhui Wu}
%\affiliation{Department of Physics, The University of Texas at Austin, Austin, Texas 78712}
\author{Alexander B. Khanikaev}
%\affiliation{Department of Physics, The University of Texas at Austin, Austin, Texas 78712}
\author{Gennady Shvets}
\email{gena@physics.utexas.edu}
\affiliation{Department of Physics, The University of Texas at Austin, Austin, Texas 78712}

\date{\today}% It is always \today, today,
\begin{abstract}
We propose a concept of a low-symmetry three-dimensional
metamaterial exhibiting a Double-Continuum Fano (DCF) optical
resonance. Such metamaterial is described as a birefringent medium
supporting a discrete ``dark'' electromagnetic state weakly
coupled to the continua of two nondegenerate ``bright'' bands of
orthogonal polarizations. It is demonstrated that light
propagation through such DCF metamaterial can be slowed down over
a broad frequency range when the medium parameters (e.g. frequency
of the ``dark'' mode) are adiabatically changed along the optical
path. Using a specific metamaterial implementation, we demonstrate
that the DCF approach to slow light (SL) is superior to that of the EIT
because it enables spectrally uniform group velocity and transmission
coefficient.
\end{abstract}

%Use showkeys class option if keyword
\maketitle

The ability to slow down light to low group velocities
$v_g$ compared with the vacuum light speed $c$ while maintaining
high coupling efficiency~\cite{lene_hau_nature99} is one of the
most dramatic manifestations of controlled light manipulation in
optics. Apart from its fundamental significance, it has
long-reaching technological applications~\cite{krauss_nphot08},
including enhanced nonlinear effects due to the energy density
compression by as much as $c/v_g$; pulse delay and storage for
optical information processing~\cite{gauthier_science07}; optical
switching, and quantum optics. Most approaches to obtaining SL rely on the phenomenon of Electromagnetically Induced
Transparency (EIT)~\cite{eit_review_harris}. EIT and its analogs
have been demonstrated in several media, including
cold~\cite{lene_hau_nature99}, warm atomic
gases~\cite{budker_prl99}, and even plasmas~\cite{shvets_prl02,avitzour_prl08}.

More recently, in response to the emerging applications such as
bio-sensing, an increasing attention has shifted towards obtaining
EIT using electromagnetic
metamaterials~\cite{zhang_prl08,zheludev_prl08,giessen_nmat09}.
Metamaterials enable engineering electromagnetic resonances with
almost arbitrary frequencies and spatial symmetries. For
example, the EIT phenomenon has been emulated in
metamaterials~\cite{zhang_prl08,zheludev_prl08,giessen_nmat09} possessing two
types of resonances: a ``dark'' one, which is not directly coupled
to the incident electromagnetic field, and a ``bright'' one, which
is strongly coupled to the incident field. If the respective
frequencies of these resonances, $\omega_Q$ and $\omega_R$, are
very close to each other, they can become strongly coupled by a
slight break of the metamaterial's symmetry. 

\begin{figure}[h]
 \centering
 \includegraphics[width=.49\textwidth]{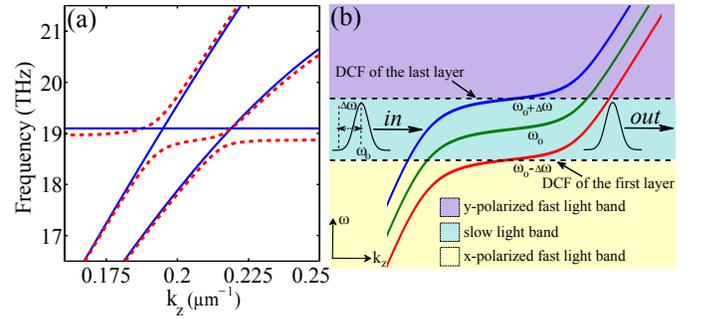} \
 \caption{(Color online) (a) PBS of DCF metamaterial as predicted
 by the analytical model of coupled oscillators [see Eqs.~(\ref{eq:oscillators},
 \ref{eq:eff_polariz})]. Different line styles correspond to the reduction of the
 spatial symmetry of a metamaterial: from two mirror-symmetry planes (solid) to
 none (dashed).(b) Broadband SL: the medium is comprised of multiple layers of
 DCF metamaterials with spatially-varying resonance frequency $\Omega_Q$ in the
 $\omega_0 - \Delta \omega < \Omega_Q < \omega_0 + \Delta \omega$ range. The incoming
 light undergoes polarization transformation and is slowed down.}\label{fig:model}
\end{figure}

The most serious limitation of the EIT-based SL (EIT-SL)
stems from the desirability of achieving the SL over a broad
spectral range, especially for optical buffering of ultra-short
pulses. Increasing coupling between the bright and dark resonances
broadens the spectral range, yet comes at the expense of
increasing both the group velocity $v_g$ and group-velocity
dispersion $d(v_g^{-1})/d\omega$. Qualitatively, the bandwidth
limitation of the EIT-SL arises because the flatness of the EIT
transmission band (which is necessary for small $d\omega/dk_z$,
where $\omega$ and $k_z$ are, respectively, the frequency and the
wavenumber of the propagating radiation) originates from the
finite spectral width $2\Delta \omega$ of the EIT band. Therefore,
only the light inside the $\omega_0 - \Delta \omega < \omega <
\omega_0 + \Delta \omega$ bandwidth can be slowed down. Using
metamaterials with spatially-dependent $\omega_0(z)$ along the
propagation direction $z$ cannot increase the total propagation
bandwidth of the EIT-SL because the EIT band is surrounded by the
stop bands as illustrated below. Therefore, a natural solution to
the bandwidth problem is to design a metamaterial supporting a
propagating mode which is {\it spectrally broad, yet possesses a
``flat'' segment} with $\partial \omega/\partial k_z \ll c$, as is
schematically shown in Fig.~\ref{fig:model}.

In this Letter we propose a new approach to producing SL
which relies on the phenomenon of Double-Continuum Fano
(DCF)~\cite{fano_pr61} optical resonance: coupling of a single
discreet state (``dark'' mode) to the two sets of continuum states
(two propagating modes of different polarization states). If the
frequency of the ``dark'' mode is embedded in the frequency
continua of the propagating modes, then a simple symmetry breaking
couples all three modes. The result of such coupling is a very
unusual propagation band [see Fig.~\ref{fig:model}(a)] which
fulfills the above requirement for broadband SL applications by
continuing from one propagating mode to another, with the SL
region in between. We demonstrate that multiple layers of such DCF
metamaterial with adiabatically changing frequency of the ``dark''
mode [as schematically explained in Fig.~\ref{fig:model}(b)] give
rise to broadband SL with spectrally uniform group
velocity and transmission. Unlike the more commonly known
single-continuum Fano resonance
\cite{kivshar_revmodphys10,lukyan_nmat10} that can be observed
with highly symmetric molecules possessing at least one
reflection mirror symmetry, DCF resonance requires low spatial
symmetry molecules that are not reflection-symmetric with respect
to any plane passing through the direction of light propagation.

Before introducing a specific metamaterial realization, we
consider a very general case of a DCF medium comprised of
anisotropic molecules containing two ``bright'' transitions
(characterized by the oscillator strengths $q_x$ and $q_y$,
respectively, coupled to the two light polarizations) and one ``dark'' transition (characterized by the oscillator
strength $Q$ and decoupled from the light). The equations for the
coupled oscillator strengths $q_x$, $q_y$, and $Q$ excited by the
light with the electric field components $E_x$ and $E_y$ are:
\begin{eqnarray}
&&\ddot{q}_x + \Omega_{Rx}^2 q_x + \kappa_{xy} q_y + \kappa_{xQ} Q = a \alpha_x E_x e^{i\omega t}, \nonumber \\
&&\ddot{q}_y + \Omega_{Ry}^2 q_y + \kappa_{xy} q_x + \kappa_{yQ} Q = a \alpha_y E_y e^{i\omega t}, \nonumber \\
&&\ddot{Q} + \Omega_{Q}^2 Q + \kappa_{xQ} q_x +\kappa_{yQ} q_y= 0,
\nonumber \\ \label{eq:oscillators}
\end{eqnarray}
where $\Omega_{Rx/Ry/Q}$ are the resonant frequencies of the
transitions and $\kappa_{xy/xQ/yQ}$ are the coupling coefficients
between them. $q_x$ and $q_y$ are coupled to external fields
through the coupling constants $a \alpha_x$ and $a \alpha_y$, where $a$ is a
constant with a dimension of $\omega^{2}$, and $\alpha_{x/y}$ are the
depolarization factors such that the dipole moment {\it normalized
to one molecule's volume $V_0$} is $p_{x/y}=\alpha_{x/y} q_{x/y}$. On
the other hand, the dark  state $Q$ cannot be directly excited by
the electric field, nor does it directly contribute to the
polarizability of the medium, thereby playing the role of the
discrete Fano state.

Solving for $q_x$ and $q_y$ in the form of $q_i = b_{ij} E_j$
(where $i$=$x$,$y$), we can construct the dielectric permittivity
tensor $\hat{\epsilon} = \hat{1} + 4\pi \hat{\chi}$, where $\hat{\chi}= (N_0
V_0) \hat{\alpha} \hat{b}$, and $\hat{\alpha}=$diag$[\alpha_x, \alpha_y]$, where $N_0$
is the molecular density. The effective polarizability of the DCF
medium is given by:
\begin{eqnarray}
&&\chi_{xx}=N_0 a \alpha_x^2 \left(W_{Ry} W_Q-\kappa_{yQ}^2\right)/D \nonumber \\
&&\chi_{yy}=N_0 a \alpha_y^2 \left(W_{Rx} W_Q-\kappa_{xQ}^2\right)/D \nonumber \\
&&\chi_{xy}=\chi_{yx}=-N_0 a \alpha_x \alpha_y \left(\kappa_{xy} W_Q-\kappa_{xQ} \kappa_{yQ}\right)/D \nonumber \\
\label{eq:eff_polariz}
\end{eqnarray}
where $D=W_{Rx} W_{Ry} W_Q-\kappa_{xQ}^2 W_{Ry}-\kappa_{yQ}^2
W_{Rx}-\kappa_{xy}^2 W_Q-2 \kappa_{xQ} \kappa_{yQ} \kappa_{xy}$ is
the determinant of the matrix formed by the l.h.s of Eq.~\eqref{eq:oscillators}, 
and $W_{i}=\left(\Omega_i^2-\omega^2\right)$. Photonic band structure (PBS) is 
analytically calculated from the eigenvalue equation for $k_z(\omega)$:
$\det(k_z^2 \delta_{ij}$$-$$\epsilon_{ij}\omega^2/c^2 )$$=$$0$,
which directly follows from the Maxwell's equations.

\begin{figure}[h]
 \centering
 \includegraphics[width=.49\textwidth]{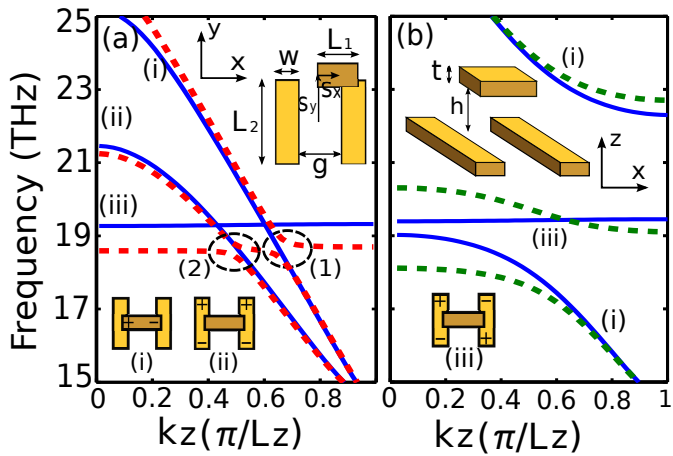} \
 \caption{(Color online) %Light propagation 
 PBS for SL metamaterials based on the DCF resonance (a) and on the EIT (b).
 The insets show the geometry and dimensions of the unit cell and
 the three supported resonances: (i) horizontal dipole, (ii) vertical
 dipole, and (iii) the quadrupole. Solid lines: PBS computed for a symmetric unit cell ($s_x=0$ and $s_y=0$). (a) Propagation bands for DCF-based
 metamaterial (dashed lines): $s_x=700$~nm, $s_y=2$~$\mu$m,
 $L_1=2$~$\mu$m. The avoided crossings marked $(1)$ and $(2)$
 are caused by $s_y \neq 0$ and $s_x \neq 0$, respectively. Flat
 portion of the spectrally-extended middle band: SL. (b)
 Propagation bands for the EIT-based metamaterials (dashed lines)
 with partial symmetry breaking ($s_x=0$~nm, $s_y=500$~nm,
 $L_1=3.8~\mu$m): emergence of SL for the spectrally-narrow
 middle band. For (a) and (b): $t=$400~nm, $h=$400~nm, $L_2=$4~$\mu$m,
 $w=$800~nm, $g=2.2$~$\mu$m. The metamaterial's periodicities are 6~$\mu$m$\times$7~$\mu$m$\times$7~$\mu$m, and the electromagnetic
waves are assumed to propagate in the $z$ direction.}
 \label{fig:band}
\end{figure}

Unlike the EIT case where the dark mode's frequency needs to be
matched to that of the $x$-dipole
resonance~\cite{zhang_prl08,giessen_nmat09}, the DCF occurs under
the following conditions: (i) $\Omega_Q < \Omega_{Rx},
\Omega_{Ry}$, and (ii) at least 2 of the 3 coupling constants
($\kappa_{xQ},\kappa_{yQ}$, and $\kappa_{xy}$) are non-vanishing.
Without significant loss of generality, and with an eye on the
specific metamaterial implementation (see schematic in
Fig.~\ref{fig:band}), we have neglected the direct coupling
between $q_x$ and $Q$. Finite values of the $\kappa$'s is the
consequence of the reduced spatial symmetry of the anisotropic
molecule.  For example, $\kappa_{xy}=0$ would be either
accidental, or due to the high spatial symmetry of the molecule
(e.g., $y-z$ mirror symmetry). Figure \ref{fig:model}(a) shows the PBS in the DCF medium (dashed line) calculated from
the analytical model using the following parameters:
$[\Omega_{Rx},\Omega_{Ry},\Omega_{Q}]=[27,23.85,19.1]$~THz,
$\left[\kappa_{xQ},\kappa_{yQ},\kappa_{xy}\right]$=$[$50,0,50$]$~THz$^2$, 
$\alpha_x$=1, $\alpha_y$=1.65, $a$=529~THz$^2$, and
$N_0V_0$=$10^{-3}$.

The unusual PBS in the DCF medium can be understood
through its evolution from that in the medium comprised of the symmetric molecules
($\kappa_{xQ}=0$, $\kappa_{xy}=0$) [solid lines in
Fig.~\ref{fig:model}(a)]. First,
the $\kappa_{xQ} \neq 0$ coupling hybridizes the $x$-dipole and the
quadrupole resonances and creates an avoided crossing between the
$x$-polarized propagation band and the discrete ``dark'' mode. The second avoided
crossing between the $y$-polarized propagation band and the discrete
``dark'' mode occurs owing to $\kappa_{xy} \neq 0$.

Thus, the DCF medium comprised of the low-symmetry molecules
supports a propagation band that smoothly connects the two
orthogonally polarized continua by passing through the SL region
near the ``dark'' mode's frequency $\Omega_Q$. Such PBS has
the following advantages over that in the EIT medium: (i) the dark mode
frequency has more tunability because it is only required to be
embedded inside the continuous bands, and (ii) there are no
bandgaps on either side of the SL region. These features of the SL
band are necessary for realizing the broadband SL using the
approach of adiabatically-varying material parameters
[specifically, $\Omega_{Q}(z)$] schematically shown in
Fig.~\ref{fig:model}(b).

Both DCF and EIT metamaterials supporting the propagation of slow
light can be implemented using a three-dimensional periodic
arrangement of the unit cells comprised of three metallic
antennas. Such unit cell shown in Fig.~\ref{fig:band} consists of
$2$ vertical and $1$ horizontal metallic antennas embedded in a
dielectric medium with $n=1.5$. It is reminiscent of the ones used
for single-layer EIT
metamaterials~\cite{zhang_prl08,giessen_nmat09}, except that both
$s_x$ and $s_y$ are allowed to be non-vanishing, thereby
dispensing with all reflection symmetries of the unit cell. The
two vertical antennas support a bright dipole and a dark quadrupole
resonances, while the single horizontal antenna supports another
bright resonance. Within the general model described by
Eqs.~(\ref{eq:oscillators}), the strengths of these resonances are
measured by $q_y$, $Q$, and $q_x$, respectively. Coupling
between all $3$ resonances is induced by breaking the reflection
symmetries of the structure. For example, $s_y \neq 0$ results in
$\kappa_{xQ} \neq 0$, while $s_x \neq 0$ results in $\kappa_{xy}
\neq 0$.

The PBS, calculated using the
finite-elements software COMSOL, is shown in Fig.~\ref{fig:band}
for different geometric parameters of the unit cell. For example,
by approximately matching $\Omega_{Rx}$ and $\Omega_Q$ (this is
done by choosing the appropriate length $L_1$ of the horizontal
antenna) and selecting $s_x=0$, $s_y \neq 0$, an EIT-SL band is
obtained as indicated by the middle dashed line in
Fig.~\ref{fig:band}(b) ($y$-polarized bands are not shown). Note
that the EIT-SL band is surrounded by the two stop-bands. 
On the other hand, when (i) $s_x \neq 0$, and (ii) the dark mode of
the fully-symmetric structure ($s_x=s_y=0$) intersects the two
propagation continua (solid lines), a DCF-SL band
emerges as shown in Fig.~\ref{fig:band}(a) (dashed lines). The
difference between the DCF bands in Figs.~\ref{fig:model}(a) and
\ref{fig:band}(a) is caused by the band-folding owing to the
periodic nature of the metamaterial in the $z$-direction.
\begin{figure}[h]
 \centering
 \includegraphics[width=.5\textwidth]{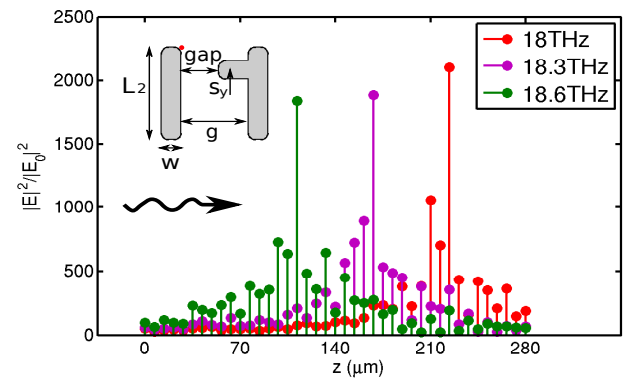}
 \caption{(Color online) Field enhancements of four different
 frequency components propagating in an adiabatically varying
 DCF-based metamaterial. Inset: single-layer unit cell. Spectral
 position of the SL is adiabatically varied:
 $L_2[\mu m] = 3.75 + z/700$. Field intensity is calculated at the
 red spot shown in the inset. Other parameters: $s_y=L_2/4$,
 $w=t=0.8$~${\mu}$m, $g=2.7$~${\mu}$m, and $gap=1.7$~${\mu}$m.}
 \label{fig:multilayer}
\end{figure}

While it is apparent from the PBS that light can be
significantly delayed by either EIT or DCF-based metamaterial, the
spectral width of such SL band is quite limited.
To create a broadband SL, we propose a multi-layered
structure with each layer having its DCF's resonance frequency
$\approx \Omega_Q$ adiabatically varied along the propagation
direction $z$. In the metamaterial structure shown in
Fig.~\ref{fig:band} such tuning can be achieved, for example, by
changing the length of the $2$ parallel antennas. Because the SL mode in the
DCF-based metamaterial contains no band gaps, the adiabatic process
smoothly converts one fast mode (e.g., $x$-polarized) into the
hybrid slow mode, and then into another fast mode ($y$-polarized)
without significant reflection. Such conversion is not possible
for EIT-SL because the SL band is surrounded by the stop bands
which lead to reflection of the SL as was recently
demonstrated~\cite{boardman_nat07,bartoli_prl08}.

Specifically, we consider a broadband $x$-polarized laser pulse with the
central frequency $\omega_0$ and bandwidth $\Delta\omega$ incident
upon an $N$-layer metamaterial with adiabatically varying
$\Omega_{Q}^{(1)} < \Omega_{Q}^{(j)} < \Omega_{Q}^{(N)}$ (where
$j$ is the metamaterial's layer number) as shown in
Fig.~\ref{fig:model}(b). It is further assumed that, while $\Delta
\omega$ is much larger than the spectral width of the SL portion
of any individual layer, the following relations are satisfied:
$\Omega_{Q}^{(1)} < \omega_0 - \Delta \omega$ and $\omega_0 +
\Delta \omega < \Omega_{Q}^{(N)}$. As the pulse passes through the
structure, each frequency component $\omega$ is slowed down inside
its corresponding layer $j$ satisfying $\Omega_{Q}^{(j)} \approx
\omega$ while propagating with no significant delay through the
other layers. The light pulse is also gradually converted from
$x$- into $y$-polarized. Because all the frequency components of the
pulse undergo the same adiabatic transition, the entire pulse is slowed down
uniformly with no group velocity dispersion.

COMSOL simulations of a $N=41$-layer DCF-SL metamaterial, with
the single-layer unit cell shown in the inset to
Fig.~\ref{fig:multilayer}, were performed. Note that this unit
cell, while topologically equivalent to the one shown in
Fig.~\ref{fig:band}, is optimized to decrease Ohmic losses.
Antennas are assumed to be made of silver with the dielectric
permittivity described by the Drude
model~\cite{Ordal_applied_optics}. The structure is
triply-periodic with the period $L=7$~${\mu}$m and embedded in a
dielectric with $n=1.5$. The SL's frequency is adiabatically
varied from layer to layer by increasing $L_2$ from
$3.75$~${\mu}$m to $4.15$~${\mu}$m. Other unit cell parameters are
given in the caption. Field intensity enhancements evaluated at
the corner of the left vertical antenna are shown in
Fig.~\ref{fig:multilayer} for $3$ different frequencies.
%The field enhancement peaks inside a specific layer indicate low group
%velocity within that layer. 
As Fig.~\ref{fig:multilayer} shows, different frequencies are slowed down inside different layers as indicating the corresponding field enhancement. Thus, light is slowed down over the entire spectral band
$\Omega_{Q}^{(1)} < \omega < \Omega_{Q}^{(N)}$.
\begin{figure}[h]
 \centering
 \includegraphics[width=.5\textwidth]{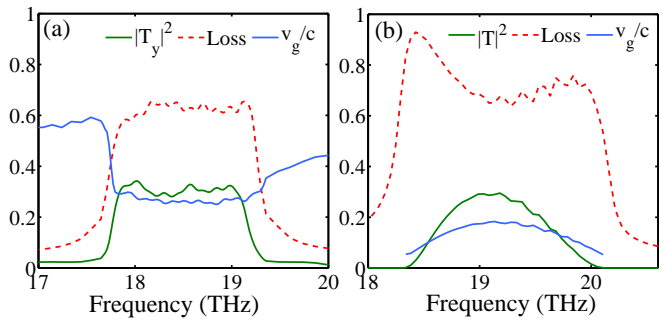} \
 \caption{(Color online) Transmission (solid green lines),
 absorption (dashed red lines), and group velocity (solid blue
 lines) of (a) adiabatic DCF-based and (b) EIT-based
 metamaterials. $x$-polarized incident light is assumed. Parameters
 of the DCF-based metamaterial: the same as in
 Fig.~\ref{fig:multilayer}. EIT-based metamaterial is made of $20$
 identical layers shown in Fig.~\ref{fig:band}, with the parameters
$L_2=4$~${\mu}$m, $L_1=3.8$~${\mu}$m, $w=t=0.8$~${\mu}$m,
$g=2.7$~${\mu}$m, $h=0.9$~${\mu}$m, $s_y=0.6$~${\mu}$m, and
periodicity $L_x=L_y=L_z=7$~${\mu}$m..}\label{fig:spectrum}
\end{figure}

To underscore the advantages of the DCF-based approach to
broadband SL, we compare a multilayer DCF-SL structure (unit
cell is shown in Fig.~\ref{fig:multilayer}) with its EIT-based
counterpart (unit cell shown in Fig.~\ref{fig:band}, geometric
parameters are given in the caption to Fig.~\ref{fig:spectrum}).
Both structures are based on silver antennas embedded in an
$n=1.5$ dielectric. Figure~\ref{fig:spectrum} (a) and (b) shows
the COMSOL simulations results for DCF-SL and EIT-SL structures,
respectively. The group velocity is calculated as $L
d\omega/d\phi$, where $L$ is the total length of the structure and
$\phi$ is the phase difference between incident and transmitted
waves. In the case of EIT, the bandwidth can be potentially
increased by increasing the coupling between dipole mode and
quadrupole mode ($\kappa_{xQ}$). However, it comes at the expense of
the increased group velocity dispersion and transmission
non-uniformity.

Specifically, both $v_g$ and transmittance $|T|^2$ increase in the
middle of the EIT band while remaining small near the EIT band
edge. In contrast, the adiabatic DCF layers provide uniform group
velocity as well as uniform cross-polarized transmittance inside
the SL band. The bandwidth can be increased further simply
by adding more adiabatically-varying layers. Note that the group velocity in the DCF-SL structure plotted in
Fig.~\ref{fig:spectrum} is averaged over $N=41$ layers. Inside the specific {\it resonant} layer, the group
velocity of the corresponding frequency component is $< 0.01c$,
which is consistent with the intensity enhancement of three orders of
magnitude. The spectrally-flat transmission and absorption of the
DCF-based metamaterial is appealing to a variety of linear and
nonlinear applications. Moreover, because every frequency
component is dramatically slowed down in a well-defined layer,
applications to spectrally-selective active light manipulation can
be envisioned.

In conclusion, we have proposed a new mechanism of slowing down
light over a spectrally broad band by means of low-symmetry
metamaterials exhibiting double-continuum Fano (DCF) resonance.
This approach is conceptually different from the more common
EIT-SL. It is shown that DCF-based broadband slow
light with uniform group velocity and transmittance can be
achieved by adiabatically changing the DCF resonance frequencies
along the light propagation direction. Spectrally-flat light
absorbers and filters, as well as various nonlinear devices
requiring extreme light concentration are enabled by DCF-based
slow-light metamaterials.

This work was supported by the grants from the Air Force Research
Laboratory and the Office of Naval Research.

\bibliography{fano_bib}
\end{document}